\documentclass[10pt, a4paper]{article}
\usepackage{lrec2022} 
\usepackage{multibib}
\newcites{languageresource}{Language Resources}
\usepackage{graphicx}
\usepackage{placeins}
\usepackage{graphics}
\usepackage{tabularx}
\usepackage{booktabs}
\usepackage{multirow}
\usepackage{flushend}
\usepackage{soul}
\usepackage{titlesec}
\titleformat{\section}{\normalfont\large\bfseries\center}{\thesection.}{1em}{}
\titleformat{\subsection}{\normalfont\SmallTitleFont\bfseries\raggedright}{\thesubsection.}{1em}{}
\titleformat{\subsubsection}{\normalfont\normalsize\bfseries\raggedright}{\thesubsubsection.}{1em}{}
\renewcommand\thesection{\arabic{section}}
\renewcommand\thesubsection{\thesection.\arabic{subsection}}
\renewcommand\thesubsubsection{\thesubsection.\arabic{subsubsection}}

\usepackage{epstopdf}
\usepackage[utf8]{inputenc}

\usepackage{hyperref}
\usepackage{xstring}

\usepackage{color}

\usepackage[colorinlistoftodos]{todonotes}

\title{KSoF: The Kassel State of Fluency Dataset -- A Therapy Centered Dataset of Stuttering}
\name{Sebastian P. Bayerl$^1$, Alexander Wolff von Gudenberg$^3$, Florian Hönig$^3$, \\ \bf \large Elmar Nöth$^2$, and Korbinian Riedhammer$^1$} 

\address{ $^1$Technische Hochschule Nürnberg Georg Simon Ohm \\
          $^2$Friedrich-Alexander Universität Erlangen-Nürnberg \\
          $^3$Institut der Kasseler Stottertherapie\\
         sebastian.bayerl@ieee.org
         }

\begin{document}

\abstract{
Stuttering is a complex speech disorder that negatively affects an individual's ability to communicate effectively. 
Persons who stutter (PWS) often suffer considerably under the condition and seek help through therapy. 
Fluency shaping is a therapy approach where PWSs learn to modify their speech to help them to overcome their stutter. 
Mastering such speech techniques takes time and practice, even after therapy. 
Shortly after therapy, success is evaluated highly, but relapse rates are high. 
To be able to monitor speech behavior over a long time, the ability to detect stuttering events and modifications in speech could help PWSs and speech pathologists to track the level of fluency. Monitoring could create the ability to intervene early by detecting lapses in fluency.
To the best of our knowledge, no public dataset is available that contains speech from people who underwent stuttering therapy that changed the style of speaking.
This work introduces the Kassel State of Fluency (KSoF), a therapy-based dataset containing over 5500 clips of PWSs. 
The clips were labeled with six stuttering-related event types: blocks, prolongations,
sound repetitions, word repetitions, interjections, and -- specific to therapy -- speech modifications.
The audio was recorded during therapy sessions at the {\em Institut der Kasseler Stottertherapie}. 
The data will be made available for research purposes upon request.
\\
\newline 
}

\maketitleabstract

\section{Introduction}

Stuttering is a complex speech disorder that affects about 1\,\% of people \cite{yairi_epidemiology_2013}.
It can be identified by an increased duration and occurrence of dysfluencies, such as repetitions, prolongations of sounds, syllables or words, and blocks while speaking \cite{lickley_disfluency_2017}. 
These so-called core symptoms are often accompanied by various linguistic, physical, behavioral, and emotional symptoms. 
Stuttering symptoms and severity vary greatly between different speakers and within the same speaker.
The unique appearance and severity of stuttering symptoms are influenced by the communication situation, psychological factors, the linguistic complexity of an utterance, and the typical phased progress of the speech disorder \cite{ellis_handbook_2009}. 
The ability to communicate can be severely disturbed and thereby negatively affect the life of a person who stutters (PWS).
Besides the high remission rate during adolescence, the condition is not curable but treatable.

There are several treatment options available that aim at different goals.
A common goal of therapy approaches is to increase the communication ability.
Some therapy approaches aim at increasing speech fluency \cite{ingham_efficacy_2015}.
In contrast, others try to make people change their attitude towards their stuttering and primarily target the psychological side-effects of stuttering \cite{mongia_management_2019}, while other approaches use a mix of the previously mentioned methods \cite{euler_computergestutzte_2009}.

Before or during therapy, the need to assess stuttering severity arises, therefore, a speech pathologist quantifies dysfluency events and types before recommending therapeutic measures. 
This is usually done during therapy sessions where PWSs perform specific speech tasks such as reading, dialogues, or scene description.
The evaluation of such tasks is highly subjective and only considers one type of communication situation, namely the therapeutic situation.
It can be shown that the use of popular evaluation metrics, such as percent stuttered syllables (\%SS), is not reliable to detect changes in one speaker if judged by only a single rater \cite{karimi_absolute_2014}.

An overall assessment in different communication situations would be ideal but laborious, and if only performed by a single therapist, would not remove subjectivity.
Realistically only the continuous automatic detection of stuttering and dysfluency symptoms inside the home of PWS or their workplace would unlock objective assessment of stuttering.
Such an assessment would enable speech therapists to plan a tailored therapy that fits the PWS's needs by using the additional information created by the assessment of the PWS speech in various communication situations.
Automatic evaluation does not only allow for a better initial assessment. 
Since stuttering therapy has high relapse rates it would benefit from speech monitoring. 
It would allow the construction of an early-warning system that enables the PWS and the speech therapist to act on the collected data and decide on further therapeutic measures. 
As many PWS have learned some speech techniques to overcome their stuttering, the usage of such would also have to be detected reliably.
\\
\\
Our contributions are:
\begin{itemize}
    \item Collection and annotation of a new therapy centered dataset containing German stuttered speech with six types of typical stuttering symptoms marked, including speech modifications.
    \item Baseline machine learning experiments for the detection of five types of stuttering as well as modified speech. 
\footnote{The annotated data will be made available to researchers upon request; please contact {\tt korbinian@ieee.org}}
    \item Insights into improving the reliability of stuttering annotations when working with naive listeners.
    \item Compatible dysfluency labels to the biggest publicly available resource containing stuttered speech, enabling cross-language transfer learning.
\end{itemize}

\section{Related Work and Data}
One of the main issues with creating reliable systems that can automatically detect stuttering in speech is too little data that captures the considerable variance in stuttering.
Most datasets are either small, not publicly available, or labeled differently, making it difficult to compare results or transfer knowledge. 

\cite{noeth_automatic_2000} used a non-public dataset of read speech with balanced classes consisting of 37 speakers and 52 recordings.
\cite{swietlicka_hierarchical_2013} used artificial neural networks (ANN) to detect three types of stuttering on a non-public dataset containing 19 speakers performing a description and a reading task. 
A much-cited resource for the automatic detection of stuttering from speech is the University College London Archive of Stuttered Speech (UCLASS) \cite{howell_university_2009}. 
\cite{kourkounakis_detecting_2020} created labels for a subset of the UCLASS corpus but did not publish the annotations. 
\cite{riad_identification_2020} used the adults who stutter (AWS) subset of the Fluency Bank corpus and created annotations for two speech and language tracks (primary and collateral), not considering blocks, prolongations, and syllable repetitions.
The LibriStutter dataset is a synthesized dataset created based on the public LibriSpeech  corpus containing labels for five types of stuttering dysfluencies \cite{kourkounakis_fluentnet_2021}.

A recent effort to solve the problem of data scarcity is the Stuttering Events in Podcasts (SEP-28k) dataset.
It consists of speech clips extracted from podcasts from and with PWS that focus on stuttering, making it by far the largest publicly available resource on stuttered speech. 
In addition to labeling the podcast data, they created compatible labels for Fluency Bank to make results easier to compare \cite{lea_sep-28k_2021,bernstein_ratner_fluency_2018}.

None of these datasets contain speech marked as using a fluency enhancing technique, as people are taught in stuttering therapy such as fluency shaping or modified phonation intervals (MPI).
\cite{swietlicka_hierarchical_2013} even explicitly asked people not to use fluency enhancing techniques therapy in their recordings. 
Reliably detecting modified speech is important, as it enables the automatic assessment of stuttering in people who already underwent stuttering therapy or use such methods in a therapeutic context. 
It allows to correctly attribute a potential gain in fluency to the speech technique.

Fluency typically improves throughout therapy \cite{euler_computergestutzte_2009,sojka_towards_2020}.
However, it is not easy to assess the level of fluency and the adoption of speech technique of a PWS after the end of therapy. 
Besides regular appointments, people are on their own, and no objective measure of therapy success is available. 
People under supervision, i.e., in a therapy environment, act or are perceived differently; therefore, the appointments can only give a snapshot of the actual performance \cite[p.~127,~205]{porta_dictionary_2014}.
Monitoring everyday adoption of speech techniques and speech fluency can provide important insight to therapists to make informed, data-driven decisions when it comes to exercises. 
It can also be used to give feedback and encourage the PWS. 

\begin{table*}[tb]
    \centering
    \begin{tabular}{l|c|c|l}
\toprule
          \textbf{Stuttering Labels} &               \textbf{KSoF} & \textbf{SEP-28k} & \textbf{Description} \\
                      
\midrule
                      Block &          20.74\,\% &   12.0\,\%  & Gasps for air or stuttered pauses \\
               Prolongation &          12.02\,\% &   10.0\,\%  & Elongated syllable or Sound “[IIII]I”, otherwi[ssss]se  \\
           Sound Repetition &          14.76\,\% &    8.3\,\%  & Repeated syllables “[nat-nat-nat-]naturally” \\
           &&& or sounds  “I [t-t-t-]talked to dad. \\
   Word / Phrase Repetition &           3.88\,\% &    9.8\,\%  &  “I have [I have] done no such thing” \\
            No dysfluencies &          24.75\,\% &   56.9\,\%  & There are no audible dysfluencies \\
 Modified/ Speech technique &          24.44\,\% &    -  \,\%  & Soft voice onset, at the start of syllables,\\
                                            &&& voluntary prolongation with continuous phonation    \\
                                            &&& e.g., rrReading, prrooolongation\\ 
               Interjection &          12.97\,\% &   21.2\,\%  &       Filler words e.g., “ähm”, “äh”, “naja”, eng: “uhm”, “uh” \\
\bottomrule

\toprule
    \textbf{Non Stuttering Labels} &                    & \\
\midrule
            Natural pause &         1.97\,\% &   8.5\,\% & A non-stuttered, significant pause in speech  \\
           Unintelligible &         2.00\,\% &   3.7\,\% &          The speech is difficult to understand \\
                   Unsure &         0.30\,\% &   0.1\,\% &         An annotator was unsure of their response \\
                No Speech &         0.39\,\% &   1.1\,\% &          The clip contains no speech or is silent \\ 
       Poor Audio Quality &         0.98\,\% &   2.1\,\% &      There are microphone or other quality issues \\
 Music (Background Noise) &         0.13\,\% &   1.1\,\% &  Audible noise or music playing in the background  \\

\bottomrule

\end{tabular}
    \caption{Distribution of annotations of 3 second segments in the Kassel State of Fluency (KSoF) dataset where at least two annotators applied a given label. SEP-28k label distribution for reference 
    \protect\cite{lea_sep-28k_2021}.}
    \label{tab:distribution}
\end{table*}

\section{Kassel State of Fluency}\label{sc:name_of_dataset}

Speech therapy has the goal to improve speech skills of people with speech and language disabilities
In the case of stuttering therapy, fluency is not necessarily the primary goal, but getting back a sense of control over one's speech.
One of the assumptions behind learning a speaking technique is that it is better to talk ``funny'' instead of not talking at all or saying something unintended. 
It takes time and effort to learn a new way of speaking that goes beyond the duration of therapy. 
At the Kasseler Stottertherapie (KST), therapy is split in three stages.
The first stage is an initial assessment and a discussion of goals with the client. 
The second stage is a two-week full-time (on-premises) intensive course in which participants learn a new speech technique and train in real-life situations, such as shopping in a bakery or flower shop or calling somebody unknown on the phone.  
The third stage follows the intensive course for one year. 
Clients use self-directed learning with the help of an online tool and occasional therapy sessions.
All recordings in this dataset were created during these three stages of therapy.

The recordings in this dataset contain three types of tasks, which are spontaneous speech (SPO), reading (REA), and telephone conversations (PHO).
SPO can be any open communication situation, like ordering at the bakery or speaking about therapy success with the therapist. 
REA is a relatively controlled task where clients were asked to read a paragraph from a given text. 
PHO is a planned but spontaneous speech task involving cold-calling unacquainted people for inquiry purposes over the phone.
For example, clients were asked to retrieve hotel booking information or opening hours and prices from a swimming pool. 
Clients can prepare in advance but have to spontaneously deal with the reactions of their dialogue partner, who is unaware that a PWS is calling.
These exercises help the PWS to grow accustomed to the speech technique and deal with possible adverse reactions of dialogue partners.  

The KSoF dataset can be used to train systems that recognize speech techniques learned to overcome stuttering.
This can help improve therapy by enabling data-driven therapeutic decisions-making that includes speech techniques.
Interventions can be made, and exercises recommended when needed and not only when scheduled.  

\subsection{Recordings}\label{ss:recordings}
This dataset consists of clips extracted from 214 recordings by 37 speakers, of which 28 were male and 9 female, containing stuttered speech. 
The language spoken throughout all clips is -- more or less regionally accented -- German.
The gender distribution matches the general ratio of males to females in stutterers, where about four times as many males stutter.
The recordings were created using either a voice recorder with a close-talking microphone or the audio was extracted from video recordings created at the initial therapy sessions.
The audio data was downsampled to 16~kHz and converted to one channel. 
The data was anonymized by removing mentions of individuals' names.
These sections in the audio signal were set to zero to ensure the participants' privacy.

\subsection{Annotation}

The recordings described in section \ref{ss:recordings} had originally been annotated with an event-based approach, while also marking the exact time spans from beginning to the end of dysfluency events \cite{valente_event-_2015,sojka_towards_2020}. 
The annotation was performed by speech therapists that had experience with stuttering therapy and the speech technique taught at KST.
Unfortunately, such accurate event-based annotation proved impractical to label large amounts of data with multiple annotators since it is time consuming and requires experienced speech therapists to perform the annotation.
The huge amount of English labeled stuttering data available through the work done by \cite{lea_sep-28k_2021} motivated us to employ a similar time-interval based annotation approach \cite{valente_event-_2015}.
A segment length of 3 seconds seems a reasonable compromise between a satisfactory level of agreement and the structure of dialogue and spontaneous speech \cite{cordes_time-interval_1994}. 

Our approach primarily differs regarding to annotator training and the annotation tool used.
The changes implemented were supposed to lead to a better agreement among annotators and thus to a better quality of the resulting labels while keeping compatibility to the large corpus to enable easy transfer learning.

Before starting the annotation process described in this paper, the recordings were manually segmented to utterances while excluding back-channels or answers of dialogue partners. 
These manually generated segments were then automatically split into 3-second long segments with 1.5-second overlap, resulting in 5597 clips. 

All clips were annotated by three annotators.
The annotators were graduate and undergraduate students from non-speech and non-health-related studies.
Prior to this task, non of the annotators had previous experience with labeling data for machine learning and had no previous prolonged exposure to PWS.
They can therefore be assumed to be naive listeners. 

The annotators were given a short 30-minute introduction to stuttering, stuttering therapy, and the labeling tool.
In addition to the introduction, written labeling guidelines that included listening examples were provided to them.
Together with the listening examples, the guidelines were accessible to them during the annotation process.
Annotators were asked to mark all stuttering- and non-stuttering-labels they could identify in the recording. 
All labels were designed as binary choices. 
Label types, the resulting label distribution, and a short description of the label can be found in Table \ref{tab:distribution}.

\begin{figure*}[htb]
\begin{center}
\includegraphics[scale=0.3]{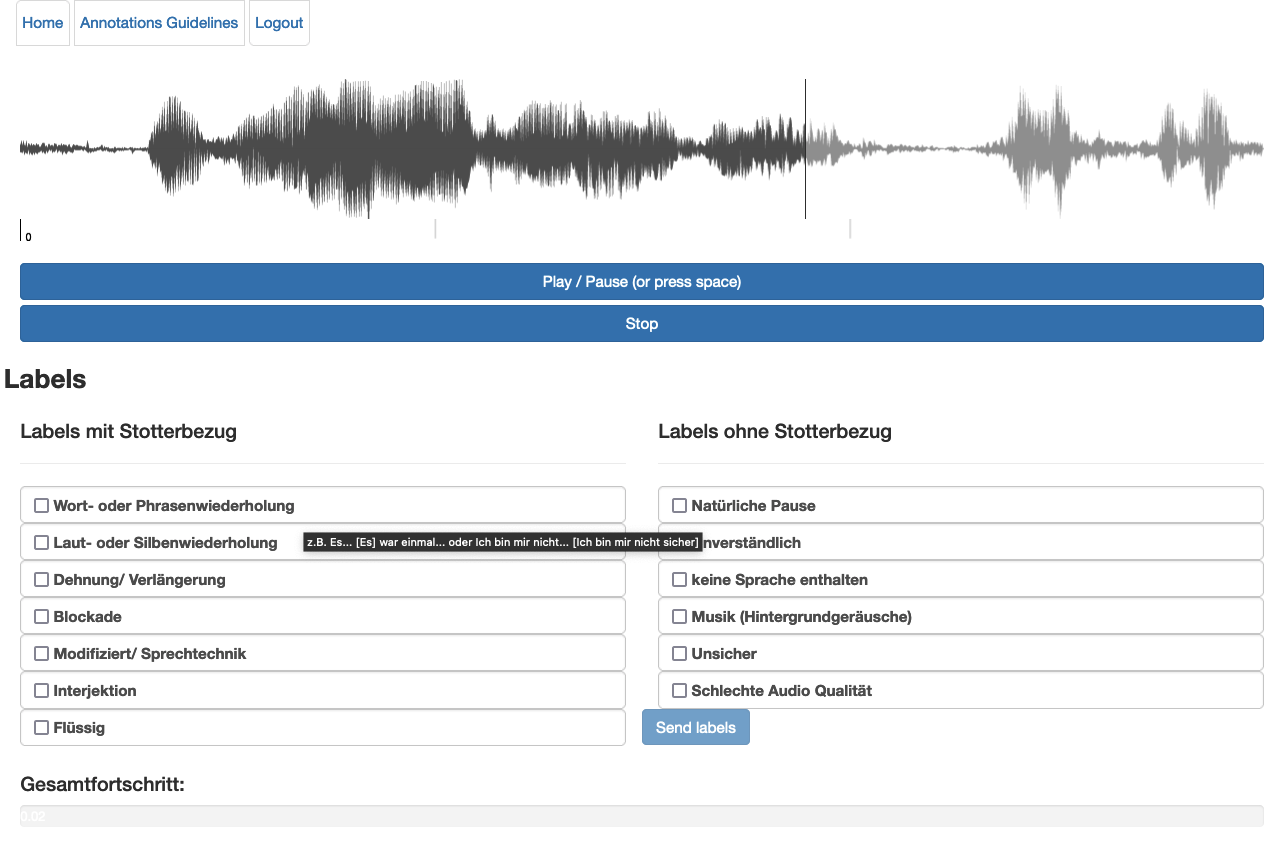} 
\caption{Screenshot of the annotation tool displaying an audio player with a plot of the audio wave-form that can be used for navigating the audio. Labels were arranged in two columns: stuttering related labels and non-stuttering related labels. A tooltip describing the label is being displayed upon hovering the mouse over the label.}
\label{fig:annotation_tool}
\end{center}
\end{figure*}

\figurename~\ref{fig:annotation_tool} shows a screenshot of the browser-based online annotation tool.
The interface featured a large audio player at the top that displays the waveform and can be used for navigation in the recording, and always displays the current location of the audio being played. 
Buttons for starting, pausing, and stopping the recording are located below the waveform plot.
Annotators could listen to the sample as many times as needed.
A tooltip was displayed when hovering over one of the label options to make the annotation task easier.
After submitting the annotations for a clip, a new clip was randomly sampled from the remaining unlabeled clips. 

The annotation process took place in two stages: the test and the main stages. 
The test stage started with a short labeling task to check the agreement between the annotators and see if there was a common understanding or misconceptions about the annotation task itself or the labels.
Annotators had to annotate 123 clips from four subjects one female and three male.
After the initial annotation task, the test stage concluded with a one-hour-long meeting with all annotators.
During the meeting, samples with a low agreement and prominent features for specific stuttering symptoms were discussed.
In the main stage of the annotation process, annotators had to assign labels to all clips in the dataset. 
They were on their own, did not communicate with each other, and only had access to the written annotation guidelines and the listening examples. 


\tablename~\ref{tab:kappa} contains Fleiss' kappa agreement metrics for all stuttering-related label types. 
The initial agreement for prolongations was very poor and improved to a moderate agreement in the main task \cite{landis_measurement_1977}.
To our surprise, interjections had the second-lowest agreement in the test stage and improved to a substantial agreement.
The agreement for no dysfluency, word repetition, and sound repetition improved slightly, whereas the value for modified stayed the same.

Overall agreement is higher than expected which can be seen by comparing values 
from agreement values achieved in a similar annotation task which can be seen in Table \ref{tab:kappa}. 

It can be hypothesized that the resulting higher-than-expected agreement was caused by the spoken instructions and introduction to the topic instead of just written guidelines and also the building of a shared understanding during the agreement meeting at the end of the test stage. 
The general agreement in the test stage and the only fair agreement for blocks in the main stage underline the difficulty of the annotation task.


\begin{table}[]
\scalebox{0.90}{
\begin{tabular}{l|c|c|c}
\toprule
            Stuttering Labels & KSoF  & Test & Sep28k   \\
\midrule
                      Block   &  0.37 & 0.60 &  0.25    \\
               Prolongation   &  0.42 & 0.06 &  0.11    \\
           Sound Repetition   &  0.54 & 0.52 &  0.40    \\
   Word / Phrase Repetition   &  0.59 & 0.57 &  0.62    \\
            No dysfluencies   &  0.59 & 0.40 &  0.39    \\
               Interjection   &  0.78 & 0.23 &  0.57    \\
 Modified/ Speech technique   &  0.55 & 0.55 &  -       \\
\bottomrule
\end{tabular}
}
\caption{Fleiss' kappa agreement statistics for each type of stuttering. Table contains values for the test labeling task (Test), the overall task (KSoF); SEP-28k agreement for comparison \protect\cite{lea_sep-28k_2021}.}
\label{tab:kappa}
\end{table}

\subsection{Metadata}
Metadata complement the annotations.
They can help with error analyses and the creation of new experiments and views of the data.   
For each clip, we provide information about
gender,
therapy status,
type of microphone used,
task performed by the PWS, 
original unique recording the clip was extracted from, 
and speaker.

\subsection{Suggested Evaluation}
Evaluation of health conditions in small datasets generally comes with some challenges. 
Specifically for stuttering, there is a considerable inter- and intra-{\em speaker} variance of stuttering behavior that is dependent of factors such as the communication situation, psychological factors, and the linguistic complexity of an utterance \cite{ellis_handbook_2009}. 
A small dataset can hardly capture this variance. 

Providing no fixed data partitioning can lead to cherry-picking and, hence, to overly optimistic results that are not reproducible, leading to an unrealistic view of the transferability of results. 
The best possible evaluation would therefore use a leave-one-speaker out approach.
For KSoF, this would require training 37 models with every experiment, which seems impractical, especially when training times are long and resources are limited. 

For small datasets such as KSoF, we believe that a speaker disjoint k-fold cross-validation (CV) is a good compromise to ensure generalization and objective results.
This evaluation strategy has its pitfalls, such as improper class- or gender distribution among folds.
Even the distribution of the communication situations might influence on fold-performance.

At the same time, we recognize the community's need for a simple and easily comparable data split into training, validation and test set.
The dataset therefore includes a suggested split for easy and quick comparison.

For KSoF, we strongly recommend using at least five-fold cross-validation when working with the data.

\section{Methods}
In the following, we briefly introduce the methods used to compute a variety of baseline classification experiments.

\subsection{openSMILE}
As our initial baseline, we chose to use the openSMILE toolkit to extract the ComParE 2016 feature set consisting of 6373 static features from the computation of various functionals over low-level descriptor (LLD) contours \cite{schuller_interspeech_2016}. 
OpenSMILE features are widely used and have been shown to achieve proper baseline performance in numerous paralinguistic applications such as gender detection, age detection, or speech emotion recognition \cite{schuller_interspeech_2016,schuller_interspeech_2021}.

We trained a Support Vector Machine (SVM) with a Gaussian kernel on the openSMILE features. 
Before training the SVM, we performed a principal component analysis (PCA) to reduce the negative effect of highly correlated features on most classification systems.
We transformed each openSMILE feature vector to a 100-dimensional vector per clip.
\subsection{wav2vec 2.0}
Neural networks benefit from large quantities of labeled training data.
Suppose this labeled in-domain data is not available in sufficient quantities.
In this case it is common to use models trained on large amounts of related data as feature encoders, as they have learned latent representations describing many aspects of the underlying data.  
The wav2vec 2.0 (W2V2) approach learns a set of speech units from large amounts of data. 
W2V2 mainly consists of a convolutional neural network (CNN) encoder, a contextualized transformer network, and a quantization module.
It takes raw wave-files as inputs, and the CNN produces latent representations that the quantization module discretizes. 
The learned units were modeled to focus on the ``most important'' factors to represent the speech audio \cite{baevski_wav2vec_2020}.
W2V2 features have already been shown to work on several speech tasks, such as phoneme recognition, speech emotion recognition, and mispronunciation detection \cite{baevski_wav2vec_2020,pepino_emotion_2021,xu_explore_2021}. 
As with dysfluencies, the ``most important'' parts of speech are disturbed.
We therefore hypothesize that these features are suitable for dysfluency detection.
We use a model pre-trained on 960 hours of unlabeled speech from the LibriSpeech corpus \cite{panayotov_librispeech_2015}.
The model was subsequently fine-tuned for automatic speech recognition (ASR) on the transcripts of the same data. 
The weights of this model were published by the authors of W2V2 \cite{baevski_wav2vec_2020}.

We extract W2V2 vectors for each audio sample for our experiments, yielding a 768-dimensional feature vector for every 20ms of raw audio. 
The W2V2 models allow the extraction of features at different layers in the feature encoder. 
Each layer yields a different representation that might be more or less suitable for a task than a later or previous layer \cite{baevski_unsupervised_2021}. 
The selected extraction layer is a tunable hyperparameter in this setup.
We then take the mean of all vectors per clip and use this as the input to train an SVM classifier with a Gaussian kernel.

\subsection{LSTM and LSTM-Attention classifier}
For our baseline LSTM-model for stutter detection we follow the baseline single target learning approach described in \cite{lea_sep-28k_2021}.
The two networks described here consist of a single layer long short-term memory (LSTM) layer with a hidden size of 64 neurons.
The last hidden state is fed into a fully connected layer for classification. 

The LSTM-Attention classifier (LSTM-A) is an extension of this model. 
An attention module complements the model with one attention head \cite{vaswani_attention_2017}.
The attention module takes all hidden states of the LSTM module as inputs with respect to the last hidden state.
The output of the attention module is then fed into the fully connected layer for classification instead of the last hidden state of the LSTM module.  
Both networks were trained with a single weighted cross-entropy loss term, using class weights. 

Both models were trained with a batch size of 64 and an initial learning rate of 0.001 and the Adam optimizer.
Early stopping was employed based on cross-validation error.

For the transfer learning (TL) experiments, we used weights from models that were pre-trained on the SEP-28k dataset with the training parameters specified in the original paper by \cite{lea_sep-28k_2021}, who used batch size of 256 with an initial learning rate of 0.01 and the Adam optimizer.
As input features to both models, we use 40-dimensional mel-filterbank energy features with a window size of 25\,ms, a frame step of 10\,ms, and frequency cut-offs at 0 and 8000~\,\textit{Hz}.

\section{Experiments}
The main distinction between the classification methods employed is the feature or feature-encoding method used.
OpenSMILE is based on handcrafted acoustic features that explicitly model prior knowledge of speech;
W2V2 is a state-of-the-art neural feature-encoder that has learned speech units that capture the essence of speech. 
The LSTM-based models were used as a sequence encoder for the traditional, signal-processing-based spectral features as a reference to the baseline system from \cite{lea_sep-28k_2021}.

\begin{table*}[!htb]
    \centering
    \scalebox{0.87}{
    \begin{tabular}{l|l|c|c|c|c|c|c}
    \toprule
     \textbf{Features}      &   \textbf{System}  &  \textbf{Mod}  &   \textbf{Bl}&  \textbf{Int} &  \textbf{Pro} &  \textbf{Snd}  &  \textbf{Wd} \\ 
    \midrule
 - &    \textbf{Random} &                  0.096  &  0.071 &  0.029 &  0.0258 &  0.038 & 0.003 \\ 
    \midrule
\textbf{openSMILE} & \textbf{SVM} & 0.58 (0.20) &  0.40 (0.14) &  0.34 (0.07) &  0.32 (0.09) &  0.36 (0.10) & 0.05 (0.07) \\ 
    \midrule
\textbf{wav2vec 2.0} & \textbf{SVM} & \textbf{0.73} (0.05) &  \textbf{0.57} (0.11) &  \textbf{0.59} (0.08) &  \textbf{0.40 }(0.03) &  \textbf{0.43 }(0.12) & \textbf{0.17 }(0.04) \\ 
    \midrule
\multirow{4}{*}{\textbf{Mel-Filterbank}} & \textbf{LSTM} & 0.36 (0.13) &  0.25 (0.09) &  0.23 (0.05) &  0.19 (0.04) &  0.22 (0.16) & 0.10 (0.02) \\ 
&\textbf{LSTM (TL)} & 0.42 (0.10) &  0.32 (0.11) &  0.25 (0.04) &  0.22 (0.01) &  0.23 (0.10) & 0.10 (0.02)  \\ 
& \textbf{LSTM-A} & 0.52 (0.09) &  0.39 (0.10) &  0.30 (0.10) &  0.26 (0.04) &  0.16 (0.06) & 0.10 (0.04) \\ 
& \textbf{LSTM-A (TL)} & 0.53 (0.08) &  0.45 (0.12) &  0.37 (0.05) &  0.29 (0.04) &  0.26 (0.15) & 0.10 (0.02) \\ 
    \bottomrule
    \end{tabular}
    } 
    \caption{Classification results are reported in the format \textbf{mean (std)} per metric for each of the labels related to stuttering: Modified (Mod), blocks (Bl), interjections (Int), prolongations (Pro), sound repetitions (Snd), and word repetitions (Wd). 
    }
    \label{tab:results}
\vspace{-2mm}
\end{table*}

\subsection{Evaluation}
All experiments use 5-fold cross-validation. 
We split the folds by speaker so no samples from a speaker in the training fold will appear in the test fold.
We report the mean F1 score per dysfluency over the five folds and the standard deviation in brackets. 
All annotations represent a binary label, and respectively all models were trained as binary classification systems in a one-vs-all approach.  
We also included the results of a fictitious random classifier using class priors for reference purposes as a lower bound for experimental results in \tablename~\ref{tab:results}.

\subsection{Results}
OpenSMILE results indicate that low-level acoustic descriptors can capture the phonetically striking dysfluency types as well as modifications.

Even though the average of all W2V2 features vectors per clip was used for classification, thereby completely ignoring the sequential nature of the problem, the SVM utilizing W2V2 features performs best consistently over all experiments and dysfluency types.
It underlines the capability of these transformer models to capture the intricacies of speech. 

The LSTM model performs poorly, which might be due to the complexity of the tasks. 
The LSTM-A model performs consistently better than the LSTM model, which is to be expected.
The attention mechanism helps the model focus on the more relevant parts of the sequence by emphasizing these inputs. 
Still, both models perform below expectations, which might be due to too little training data.
The initial cross-language transfer-learning experiments increase performance over all dysfluency types but word repetitions for both the LSTM and LSTM-A system.
These results support this assumption.

Modifications are the class that is consistently detected best regardless of model. 
This is not surprising as it is a very distinctive pattern and a learned behaviour shared among all PWS recorded for this dataset, probably making it the most straightforward pattern to detect.


Performance of all approaches is consistently worst detecting word repetition.
One factor is the small number of positive examples, only 3.8~\% of clips in the corpus, which makes training difficult. 
Another factor is that of all the types of stuttering, word repetitions need the longest context to be recognizable.
At the same time, they are acoustically almost indistinguishable from non-dysfluent speech, as they are just a repetition of a word.
The pattern is complex as it spans large parts of a clip.
It could probably be detected most reliably with an ASR system and an adapted language model, as suggested by \cite{alharbi_automatic_2017}.

\section{Discussion and Outlook}
We presented KSoF, a new resource of stuttered speech including speech of PWS who learned to modify their speech.
This unique resource will enable more research into the automatic assessment of stuttering severity in a therapeutic context. 
The rather simple labeling approach proves to be reliable and cost-effective by utilizing naive annotators.
It could be shown that even little training can help the inter-rater agreement and consequently also the reliability of labels. 

Baseline machine learning experiments -- while promising -- raise questions to be addressed by future research. 
It is unclear why the spectral features and the LSTM system performed below expectations, especially when comparing results to the hand-crafted heuristic features extracted with openSMILE.
A detailed error analysis per feature and dysfluency type could shed light on this. 
Prolongations are the dysfluency type that is phonetically most similar to modified speech, which might be a factor for the rather poor baseline performance.
A detailed analysis of misclassified clips can help to answer these questions. 
A detailed look at how W2V2 encodes the audio might also reveal why the predictive power of those features is so big, even when averaging values over whole clips. 

For future work, we plan to explore multi-class classification of stuttering with a single classifier for all types of stuttering.
Future research should focus on detecting and localizing of dysfluency events in continuous speech.
This can help to improve ASR systems by identifying people with dysfluent speech, and speech therapy applications can profit from precise automated feedback. 
We also encourage researchers to explore different aspects of the data, such as the recording situation that might lead to alternative experimental settings.

\subsection*{Acknowledgment}
We thank Jo\"{e}lle Döring for providing the initial annotation and segmentation and the rewarding discussions of all aspects of fluency and dysfluency. 
This work was supported by the Bayerisches Wissenschaftsforum (BayWISS).

\section{Bibliographical References}\label{reference}

\bibliographystyle{lrec2022-bib}
\bibliography{references}

\end{document}